\documentclass[prl,showpacs,preprintnumbers,twocolumn,superscriptaddress,floatfix]{revtex4-1}
\usepackage[utf8]{inputenc}
\usepackage{bm,amsmath,amssymb}
\usepackage{graphicx}
\usepackage{subfigure}
\usepackage{color}
\usepackage{hyperref}
\usepackage{multirow}
\usepackage{soul}
\usepackage{comment}
\let\oldalign\align
\def\align{\linenomath\oldalign}
\newcommand{\xs}{{\bf x}}

\begin{document}
%\linenumbers

\title{Understanding the approach to thermalization from the eigenspectrum of 
non-Abelian gauge theories  }

\author{Harshit Pandey}
%\email{harshitp@imsc.res.in}
\affiliation{The Institute of Mathematical Sciences, a CI of Homi Bhabha National Institute, Chennai 600113, India}
\author{Ravi Shanker}
%\email{rshanker@imsc.res.in}
\affiliation{The Institute of Mathematical Sciences, a CI of Homi Bhabha National Institute, Chennai 600113, India}
\author{Sayantan Sharma}
%\email{sayantans@imsc.res.in}
\affiliation{The Institute of Mathematical Sciences, a CI of Homi Bhabha National Institute, Chennai 600113, India}

%\date{\today}
%------------------------------------------------------------------------------------
%       abstract
%------------------------------------------------------------------------------------
\begin{abstract}
We study some interesting aspects of the spectral properties of SU(3) gauge theory, both with and without 
dynamical quarks (QCD) at thermal equilibrium using lattice gauge theory techniques. By calculating the 
eigenstates of a massless overlap Dirac operator on the gauge configurations, we implement a gauge-invariant 
method to study spectral properties of non-Abelian gauge theories. We have unambiguously categorized Dirac eigenvalues 
into different regimes based on a quantity defined in terms of the ratios of nearest neighbor spacings. While majority 
of these eigenstates below the magnetic scale are similar to those of random matrices belonging to the Gaussian Unitary 
ensemble at temperatures much higher than the chiral crossover transition in QCD, a few among them start to 
become prominent only near the crossover. These form fractal-like clusters with the median value for their fractal 
dimensions hinting at the universality class of the chiral transition in QCD. We further demonstrate that momentum 
modes below the magnetic scale in a particular non-equilibrium state of QCD are classically chaotic and estimate 
an upper bound on the thermalization time $\sim 1.44$ fm/c by matching this magnetic scale with that of a thermal 
state at $\sim 600$ MeV.

\end{abstract}

\maketitle

%-----------------------------------------------------------------------------------------------------
%-----------------------------------------------------------------------------------------------------

\textbf{Introduction}
The eigenspectrum of the Hamiltonian of an isolated quantum system provides valuable 
insights not only on dynamical properties such as their approach to thermalization but 
also their properties in thermal equilibrium through the eigenstate thermalization hypothesis 
(ETH)~\cite{PhysRevA.43.2046,Srednicki:1994mfb,Tasaki:1998yvw,Rigol:2007juv}. 
Bohigas, Giannoni and Schmit (BGS) conjectured that~\cite{Bohigas:1983er} quantum fluctuations 
of the eigen level-spacing distributions whose classical dynamics is chaotic can be described, 
based on symmetries, by one of three distinct Wigner classifications of random matrix theory (RMT). 
Whereas this conjecture has been verified in spin systems with different global as well as local 
symmetries but no such verification exists yet for non-Abelian gauge theories. Generic 
quantum systems, however, have a much richer spectrum owing to the melding of regular levels with 
the chaotic ones~\cite{MVBerry_1984} due to which the realization of BGS conjecture is 
non-trivial~\cite{Bohigas:1990zz}. Furthermore, eigenstates with universal 
RMT fluctuations may provide a route towards thermalization in (strongly) interacting isolated 
quantum systems~\cite{polkovnikov2011colloquium,annurev:/content/journals/10.1146/annurev-conmatphys-031214-014726}.

Notwithstanding the challenge to construct a quantum Hamiltonian for non-Abelian gauge theories like QCD,
studying its spectral properties is important to understand how such systems with strong color (gauge) 
interactions thermalize.  This is relevant for addressing questions like, in what time scale 
will the fireball formed in a relativistic heavy-ion collision or the quarks and gluons 
in the early universe thermalize~\cite{Schlichting:2019abc,Berges:2020fwq}. In the latter case, 
this is crucial for understanding nucleosynthesis of light elements that accounts for $\sim 99\%$ 
of the visible matter in the present universe. It will also be important to investigate, once thermalized, 
if spectral properties of QCD have some imprints of the universal features related to chiral symmetry 
restoration at QCD phase transition, even though the universal critical part of thermodynamic quantities 
like free energy may be subdominant compared to its regular part for physical quark mass.

From ab-initio lattice field theory simulations it is now established that thermal non-Abelian 
SU(3) gauge theory coupled to two dynamical light quarks and a heavier strange quark that transform 
under the fundamental representation of the gauge group undergoes a smooth crossover transition~
\cite{Aoki:2006we,Bazavov:2011nk,HotQCD:2014kol,Bhattacharya:2014ara,Burger:2018fvb}. This is
due to the fact that QCD with two light quarks has an almost exact non-singlet $SU_V(2)\times SU_A(2)$ 
chiral symmetry which breaks into its $SU_V(2)$ subgroup below a pseudo-critical 
temperature $T_c=156.5(1.5)$ MeV~\cite{HotQCD:2018pds}. The confinement of color degrees 
of freedom also happens around the same temperature~\cite{Bazavov:2016uvm}, leading to 
the formation of color-singlet baryon and meson states. The singlet $U_A(1)$ part of  
chiral symmetry, though anomalous, is believed to affect the nature of the phase transition in
QCD with two light quark~\cite{Pisarski:1983ms} flavors depending on its \emph{effective} 
magnitude at $T_c$, and is an ongoing area of research~
\cite{Aoki:2012yj,HotQCD:2012vvd,Tomiya:2016jwr,Dick:2015twa,Brandt:2016daq,Petreczky:2016vrs,Aoki:2020noz, Ding:2020xlj,Aoki:2021qws, Kaczmarek:2021ser,Kaczmarek:2023bxb,Kovacs:2023vzi,Alexandru:2024tel,Giordano:2024jnc}. 
Specific features of the Euclidean Dirac eigenvalue spectrum in QCD can successfully explain the 
breaking or restoration of the non-singlet chiral symmetry~\cite{Smilga:1993in,Verbaarschot:1993pm}. 
In absence of any conclusive evidence on whether the $U_A(1)$ is restored \emph{effectively} with its 
non-singlet counterpart at the same temperature, one might ask if any specific features of the eigenspectrum 
in QCD can provide a fresh perspective on this matter. 

In this Letter, we motivate and show how some specific features of the QCD Dirac eigenspectrum 
can give us important insights to address these unresolved issues. In the Dirac spectrum, the localized and 
the chaotic regions can be separated out which allows us to provide a first demonstration of the realization of BGS 
conjecture in a non-Abelian gauge theory in three spatial dimensions using lattice techniques. For this, we 
show using some discerning observables that the eigenstates of the Dirac operator in 2+1 flavor QCD 
with eigenvalues below the magnetic scale~\cite{Linde:1980ts} at a temperature $T\simeq 624 $ MeV, have spectral 
properties consistent with a particular RMT universality. There also exists a particular non-equilibrium 
state where the low-momentum single-particle gluon states below the corresponding magnetic scale are densely occupied, 
hence classical~\cite{Lappi:2006fp,Gelis:2010nm,Berges:2013fga} and is chaotic, characterized by 
a positive Lyapunov exponent. A crucial observation from our study is that both in the classical non-thermal 
state as well as in the quantum (thermal) state of QCD at high temperatures, these single-particle states of gluons 
below the non-perturbative \emph{magnetic} scale~\cite{Linde:1980ts} have the same physical properties akin to a 
chaotic system. Matching the scales that define these states enables us to provide an upper bound for 
the thermalization time $\simeq 1.44$ fm/c to arrive at the thermal state described by $T \simeq 624$ MeV, starting from 
the over-occupied non-thermal state.

The presence of light dynamical fermions (quarks) does not affect our procedure of matching scales since 
the chaoticity inherent in the spectrum, evident from the universal fluctuations akin to RMT, is a property of the non-Abelian 
color group. Rather, their presence only contributes to the magnitude of the \emph{magnetic} scale through the running of the 
coupling, thus ensuring a faster thermalization of the magnetic gluons.  Once thermalized, the role of dynamical quarks becomes 
more important close to the chiral crossover transition. We study how the spectral properties of the thermal states change as the 
temperature is cooled down to $T_c$ where the presence of dynamical light quarks indeed becomes increasingly relevant in determining 
the structural properties of the infrared eigenstates and thus driving the chiral crossover transition. The properties of the 
eigenspectrum of QCD Dirac operator on thermal states of QCD as a function of temperature are discussed in the third section after 
briefly describing our numerical setup. The subsequent sections discuss the properties of a particular non-thermal state of QCD thus 
building upon the discussion on estimating an upper bound on the thermalization time for the magnetic gluons in QCD. We conclude by 
discussing the implications of our study.

\textbf{Numerical Set-up}
The thermal partition function of QCD, written in terms of its quantum Hamiltonian 
$H_{\text{QCD}}$ as $\text{Tr [e}^{-H_{\text{QCD}}/T}]$, can be equivalently represented 
as a path integral in the configuration space of fermions and gauge fields, weights 
of which are determined by $\rm{e}^{-S_{\text{QCD}}}$, 
where $S_{\text{QCD}}$ is the classical action in Euclidean spacetime with a compact 
\emph{fictitious} time direction $\tau$ of extent $1/T$. If a probe massless quark is in 
thermal equilibrium with these gauge fields, the information encoded in the eigenspectrum 
of the QCD Hamiltonian, e.g., the universal level spacing fluctuations,  will be contained 
in its Dirac eigenspectrum as well~\cite{Giordano:2021qav}.

In order to study the Dirac eigenspectrum of QCD with two light and one 
heavier strange quark, we have used gauge configurations generated using the M\"{o}bius 
domain wall discretization for fermions~\cite{Kaplan:1992bt} which respects chiral 
symmetry on the lattice to a very good extent. This particular domain wall 
fermion discretization allowed us to work with a moderately small fifth-dimensional extent,  
\( L_5 = 16 \) and the M\"{o}bius parameters \( b_5 \) and \( c_5 \) 
were chosen such that \( b_5 - c_5 = 1 \). The values of \( c_5 \) at different temperatures 
are listed in Table~\ref{tab:table1}, which were chosen such that the chiral symmetry violation 
for our chosen $L_5$ is minimal. The domain wall height \( M_5 \), was kept fixed at $1.8$ for 
the entire set of simulations. The bare light and strange quark masses were 
chosen such that the pion mass is physical i.e. $M_\pi \sim 135$ MeV and so is the kaon mass 
with $M_K \sim 495 $ MeV~\cite{Bhattacharya:2014ara}.  For $T\lesssim 186$ MeV we 
have partly used the existing gauge configurations generated by the HotQCD
collaboration~\cite{Bhattacharya:2014ara} and a few from Ref.~\cite{Gavai:2024mcj} 
whereas the configurations for $T=195$ MeV were generated by us choosing the same set 
of parameters used in Ref.~\cite{Bhattacharya:2014ara}. 

We have worked with a fixed lattice size with $N_s=32$ sites along each spatial direction 
and $N_\tau=8$ sites along the Euclidean temporal direction. The range of temperatures, 
$T=1/(N_\tau.a)$, we have studied is between $149$-$195$ MeV, where $a$ is the lattice 
spacing, such that the spatial extent $L=V^{1/3}$ is$\sim 5.3$ fm for $T=149$ MeV whereas 
$L\sim 4$ fm for the highest temperature $T=195$ MeV. We use the massless overlap Dirac 
operator~\cite{Narayanan:1994gw,Neuberger:1997fp} defined as 
$D_{ov} = 1+ \gamma_5~\textit{sign}(\gamma_5 D_w(-M_5))$ where $D_w(-M_5)$ is the Wilson Dirac operator 
with the domain wall height $M_5 = 1.8$ (appearing as a negative mass parameter) as a probe to measure 
the eigenspectrum of the QCD configurations. This is because the overlap Dirac operator has an exact chiral 
symmetry on the lattice. For the implementation of the matrix-sign function, we have calculated the lowest 
$30$ eigenvectors of $D_w^\dagger D_w$. The sign function was computed exactly in the vector space 
spanned by first $30$ eigenvectors of the Wilson Dirac operator and approximated in terms of a Zolotarev rational 
polynomial with $25$ terms in the rest of the vector space. The numerical precision of the matrix sign-function was 
such that its norm-squared deviated from unity at $\sim 10^{-9}$-$10^{-10}$ and the resultant overlap Dirac operator satisfied the 
Ginsparg-Wilson relation at a precision $\sim 10^{-10}$. We have calculated the first $\sim 100$ eigenvalues of the 
overlap Dirac operator using the Kalkreuter-Simma Ritz algorithm~\cite{Kalkreuter:1995mm}. We have also generated 
thermalized SU(3) gauge configurations (without dynamical quarks) with Wilson gauge action on $32^3\times 8$ 
lattice and inverse coupling $\beta=6/g^2=6.545$, which corresponds to a temperature $T \simeq 624$ MeV, 
about twice the deconfinement temperature $T_d\simeq 312$ MeV~\cite{Boyd:1995zg,Francis:2015lha}. The scale 
$r_0=0.4701(36)$ fm~\cite{FlavourLatticeAveragingGroupFLAG:2021npn} is used to set the temperature in this case. At this 
temperature the gluon degrees of freedom, especially below the magnetic scale are dominant, thus these gauge ensembles 
represent the physical scenario. The details about the number of eigenvalues computed and the number of configurations 
used at each temperature in this study are summarized in Table~\ref {tab:table1}.

The SU(3) gauge configurations at high temperatures were generated in order to compare how different 
a thermalized state is from a classical state of QCD which consists of highly occupied gauge 
fields~\cite{Berges:2013eia,Berges:2013fga} with a non-thermal phase space distribution and without 
dynamical quarks. For generating these non-thermal, classical gauge configurations on a three-dimensional 
lattice using Hamiltonian evolution, we choose the lattice extent with $N=64$ sites along each spatial 
direction, with lattice spacing in units of the gluon saturation scale $Q\sim 1.5$ GeV to be $Qa_s=0.25$ 
such that the spatial extent in physical units is $\sim 2.1$ fm. This is to ensure that we 
have a sufficiently large volume and enough number of momentum modes of gauge fields
in the deep-infrared part of its momentum distribution.

\begin{table}[ht]
 \centering
 \begin{tabular}{|c|r|r|r|r|r|c|}
 \hline \hline
 2+1 flavor QCD & & & & & & \\
 \hline
 \hline
$T$ (MeV) & $\beta$ & $c_5$&$N_s$ & $N_\tau$  & $N_{\text{eigen}}$ & $N_{confs}$ \\
 \hline
  149 & 1.671 & 1.5 &32 & 8 &100 &40 \\
  159 & 1.707 & 1.5&32 & 8 &100 & 44  \\
  164 & 1.725 & 1.5 &32 & 8 &100 & 95 \\
  177 & 1.771 &1.0 &32 & 8 &100 & 93 \\
  186 & 1.801 &1.0 &32 & 8 &90 & 57 \\
  195 & 1.829 &0.9 &32 & 8 &100 &101 \\
  \hline
 \hline
 SU(3) & & &  & & & \\
 \hline
 \hline
 624 & 6.545 &---& 32 & 8 &100 & 46 \\
\hline
  \end{tabular}
  \caption{Number of gauge configurations studied at different temperatures (corresponding to the values of \( \beta =6/g^2\)) and the lattice sizes used in this work. The table also lists the number of overlap Dirac eigenvalues calculated per gauge configuration, along with the values of the M\"{o}bius parameter \( c_5 \) used in the generation of gauge ensembles at each temperature. }
  \label{tab:table1}
\end{table}

\textbf{Properties of eigenvalue spectrum of gauge theories in thermal equilibrium}
We study the spectral properties of the four-dimensional overlap Dirac operator of a probe 
massless quark whose eigenvalues are denoted as $\lambda_n$ where $n$ is an integer whose 
maximum value is the dimension of the Dirac operator.  This is a gauge-invariant procedure 
to understand the spectral properties of QCD non-perturbatively. The fluctuations of level 
spacings between consecutive eigenvalues are universal and belong to one of the three Wigner surmises
if the system is chaotic. An important observable of interest to study these fluctuations is 
$\langle \Tilde{r} \rangle $, where $\Tilde{r}_n $ is defined~\cite{PhysRevB.75.155111} as 
$  \Tilde{r}_n = \text{min}\left( r_n,\frac{1}{r_n}\right)$. The quantity $r_n = \frac{s_{n+1}}{s_n}$ 
represents the ratio between two consecutive level spacings $s_n = \lambda_{n+1}- \lambda_n~ $. 
The observable \( \langle \tilde{r} \rangle \) is obtained by averaging over \( \tilde{r}_n \) 
for all eigenvalues within a spectral window and subsequently over the entire ensemble
of gauge configurations at a particular temperature.  For this reason, the index \(n\) is 
explicitly dropped after performing these averages while calculating $\langle \Tilde{r} \rangle$. 
There is a distinct advantage of choosing $\langle \Tilde{r} \rangle $ over other 
observables discussed in the literature~\cite{Kovacs:2017uiz} to characterize the eigenvalue spectrum. 
This is due to the fact that, unlike observables which are integrated quantities of level-spacing distributions 
and thus require unfolding to remove the system-dependent mean, the quantity $\langle \Tilde{r} \rangle $ does 
not suffer from such systematic uncertainties. 

\begin{figure}[h]
    \centering
    \raisebox{-\height}{\includegraphics[width=0.49\textwidth]{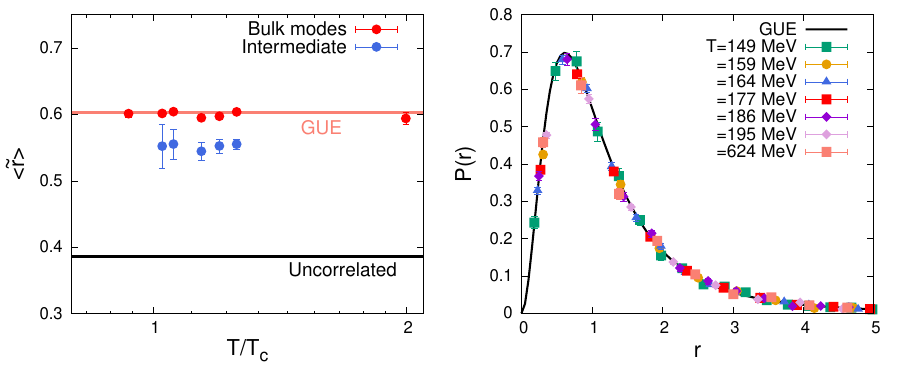}}
    \caption{Left panel: Average value \( \langle \tilde{r} \rangle \) in different spectral windows defined 
    in terms of the overlap Dirac eigenvalues at various temperatures \( T \). Red and blue points correspond to bulk and 
    intermediate eigenmodes, respectively. The results are compared with the theoretical expectations for the GUE and 
    completely uncorrelated (Poisson) spectra, shown as red and black horizontal lines, respectively. Right panel: 
    Probability distribution of ratios of     consecutive level spacings for bulk eigenmodes at different temperatures 
    with and without dynamical quarks. Note that for pure SU(3) gauge ensemble at $T=624$ MeV, the $T_c$ 
    represents the deconfinement temperature $T_d$.}
    \label{fig:average_rtilda}
\end{figure}

We now elucidate our procedure to categorize different spectral windows. We first calculate 
$\langle \Tilde{r} \rangle$ in each small bins of $\lambda/T$ by averaging over the gauge configurations at each 
temperature. We then checked if the values of $\langle \Tilde{r} \rangle$ are exactly 
similar within errors to that obtained for random matrices belonging to a Gaussian Unitary Ensemble 
(GUE) for some of the bins.  Such bins then define a particular spectral window. For $T<T_c$, and 
for $T\sim 2T_d$ the values of $\langle \Tilde{r} \rangle$ measured in all the bins  
are consistent within errors with $\langle \Tilde{r} \rangle=0.60266$ obtained for the GUE. 
We observe a similar agreement for all bins within the spectral windows defined by 
$\lambda/T>0.08, 0.12, 0.33, 0.5, 0.6 $ at temperatures $T=159, 164, 177, 186, 195$ MeV respectively. 
We note here that the upper cutoff to these spectral windows is decided by the maximum eigenvalue that 
we have calculated at a particular temperature. We henceforth refer to them as \emph{bulk eigenmodes} 
and their resulting $\langle \Tilde{r} \rangle$ as a function of $T/T_c$ are shown as red points in 
the left panel of Fig.~\ref{fig:average_rtilda}. These bulk eigenmodes thus characterize the chaotic 
part of the QCD Dirac eigenspectrum.

We next define \emph{intermediate eigenmodes} which belong to the spectral window
characterized by $\lambda/T$ between $0.0$-$0.08$ for $T=159$ MeV, $0.0$-$0.12$ for 
$T=164$ MeV, $0.0$-$ 0.33$ for $T=177$ MeV, $0.0$-$0.5$ $T=186$ MeV and $0.0$-$0.6$ 
for $T=195$ MeV, for which the $\langle \Tilde{r} \rangle$  for all the bins within the spectral window lies between the GUE prediction 
and the corresponding value for completely uncorrelated eigenvalues.  The emergence of these 
intermediate eigenmodes near $T_c$, shown as blue points in the left panel of Fig.~\ref{fig:average_rtilda}, 
has important physical consequences which we will discuss in the later part of this section.

To verify that the bulk eigenmodes are correctly identified and the robustness of our procedure,  
we next measure the probability distribution of nearest neighbor spacing ratios~\cite{CHAVDA20133009} 
since this quantity is also independent of the unfolding procedure and can be directly matched with 
the GUE prediction given by $P(r)=\frac{11.16~ (r+r^2)^2}{(1+r+r^2)^4}$. In the right panel of 
Fig.~\ref{fig:average_rtilda}, we observe that the bulk eigenmodes defined using our criterion indeed 
follow the GUE prediction for all temperatures. The spectral window that defines the bulk eigenmodes is 
similar to our earlier studies of the continuum extrapolated level spacing distribution in QCD with the 
staggered Dirac discretization~\cite{Kaczmarek:2023bxb}.

\begin{figure}[h]
    \centering
    \raisebox{-\height}{\includegraphics[width=0.49\textwidth]{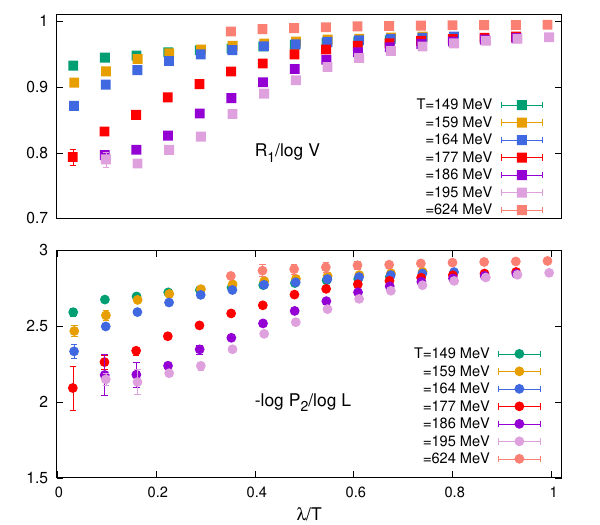} }   
    \caption{Variation of first Rényi entropy (top panel) and $D_2$ (bottom panel), 
    for eigenvectors binned in $\lambda/T$ for different temperatures, performed on the gauge configurations mentioned in table~\ref{tab:table1}. 
    }
    \label{fig:Renyi}
\end{figure}

We next study the localization properties of the Dirac eigenstates, which for a particular 
eigenstate $\psi(\xs)$  is quantified with the generalized Rényi entropy defined 
as,
\begin{equation}
R_\alpha=\frac{1}{1-\alpha} \ln \sum_\xs p_\xs^\alpha~,~1\leq \alpha<\infty~.
\end{equation}
The quantity $p_\xs= |\psi(\xs)|^2 $ denotes the probability density of an eigenstate at a 
spatial position $\xs$ obtained after averaging the four-dimensional density over $\tau$. 
From the top panel of Fig.~\ref{fig:Renyi} it is evident that the bulk eigenmodes 
have significantly higher values of the first Rényi entropy $R_1$ ranging between $0.92 < R_1 /\log V < 1.0$, 
approaching unity at the highest temperature i.e. $624$ MeV. This implies that the bulk eigenmodes are 
almost completely delocalized over the spatial volume and thus are ergodic, containing equivalent 
thermal information in each of them. This is an indirect manifestation~\cite{Rigol:2007juv} 
of the ETH, which has been demonstrated in non-Abelian theories in 2+1 dimensions~\cite{Yao:2023gnm,Ebner:2024mee}. 
Whereas close to $T_c$ at $T= 159, 164, 177, 186, 195$ MeV, the intermediate eigenmodes do not contain the entire 
thermodynamic information that is encoded in the bulk eigenmodes, as evident from their values of $R_1/\log V$. 
These intermediate eigenmodes are thus not as delocalized as bulk eigenmodes, hence not fully ergodic. We will next 
discuss the implications of this observation.

\begin{figure}[h]
    \centering
    \raisebox{-\height}{\includegraphics[width=0.50\textwidth]{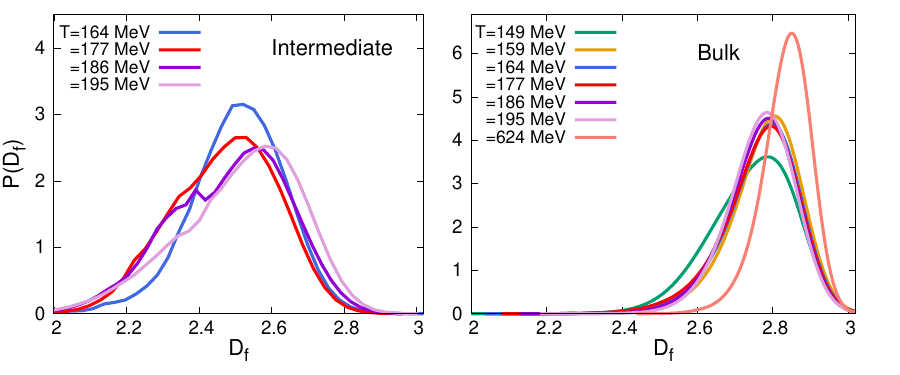}}
    \caption{  Probability distribution of the fractal dimension $D_f$ at different temperatures and in eigenvalue intervals characterized as the intermediate (left panel) and the bulk eigenmodes (right panel). }
    \label{fig:Df}
\end{figure}

To further understand the structure of the intermediate eigenmodes, we study the inverse 
participation ratio (IPR) defined as $P_2 = \sum_{\mathbf{x}}|\psi(\mathbf{x})|^{4}$. For 
a completely delocalized eigenstate we expect $P_2$ to scale as the inverse of the volume 
of the system as $L^{-3}$ but may not achieve this scaling even in the deep bulk due to the
presence of thermal fluctuations. The plot of $D_2=-\log P_2/\log L$ as a function of bins in 
$\lambda/T$ shown in the lower panel of Fig.~\ref{fig:Renyi} demonstrates different 
volume scaling for the bulk and intermediate eigenmodes. To further quantify this scaling we calculate 
the distribution of the fractal dimension, $D_f$ of the eigenvectors, in the eigenvalue intervals used 
earlier to characterize the intermediate and bulk eigenmodes instead of $D_2$, because of the low statistics. 
However both of these quantities contain equivalent information about the scaling of the eigenmodes.
We have calculated the fractal dimension of an eigenstate using the box-counting method~\cite{mandelbrot1983fractal}
where the threshold for the probability density is chosen to be $0.75/V$, since this choice yields a scaling 
behavior that is consistent with $D_2$ within errors across all $\lambda /T$ bins. The probability distribution in 
Fig.~\ref{fig:Df} of the fractal dimension $D_f$ has been constructed using an error-weighted Gaussian smearing method, 
where the data is extracted from a Gaussian distribution centered at each measured value of $D_f$ with a width given 
by its error. These data are summed and normalized such that the area under the curve is unity, thus producing a 
smooth probability distribution function instead of discrete histograms. As evident from Fig.~\ref{fig:Df},
the fractal dimensions of the bulk eigenmodes for all temperatures are significantly higher than the 
intermediate eigenmodes, the median values ranging between $2.78 \lesssim D_f\lesssim 2.84$. For the temperature range 
$\gtrsim T_c$, the $D_f$ of the intermediate eigenvectors show a sensitivity to temperature variation thus reflecting the 
physical changes occurring in the system due to the chiral crossover transition. The probability distribution of $D_f$ 
denoted by $P(D_f)$ has a spread due to the fact that fluctuations can exist on all length scales larger than the lattice 
spacing resulting in wavefunctions with different extent of delocalization to co-exist~\cite{PhysRevLett.83.4590}, 
with median values of $D_f=2.52, 2.49, 2.52, 2.55$ at $T=164, 177, 186, 195$ MeV respectively, for the intermediate 
region. The median value of $D_f$ just above the thermal crossover transition can be interpreted using the relation 
$D_f=3-\beta/\nu$~\cite{Jansen:1990jy}. If we put in the critical exponents $\beta,\nu$ of a three-dimensional O(4) 
spin model~\cite{Engels:2011km}, the value of $D_f\simeq 2.485$, which is similar to the median values of $D_f$ of the 
intermediate eigenvectors.

The presence of small kinks in the probability distribution of $D_f$ at $T\gtrsim 186$ MeV is expected due to the 
rare occurrence of near-zero modes at high temperatures, hence one would require larger statistics to obtain a 
smooth curve. One also observes a spread of $P(D_f)$ towards lower values of $D_f \simeq 2$ with increasing temperatures, 
indicating the appearance of localized eigenmodes. With this trend we expect that at even higher temperatures, localized 
eigenmodes characterized by $\langle \Tilde{r} \rangle$ close to the value in a uncorrelated Poissonian distribution would 
contribute significantly to the broadening of $P(D_f)$. This can be related to the observation of the Anderson-Mott 
transition and the existence of a mobility edge at high 
temperatures~\cite{Kovacs:2010wx,Kovacs:2012zq,Kovacs:2017uiz,Giordano:2013taa,Giordano:2016cjs, Giordano:2016nuu,Holicki:2018sms,Giordano:2021qav,Alexandru:2021xoi,Giordano:2022ghy,Kehr:2023wrs}.
Eventually the low-lying eigenstates near or below the mobility edge, will exhibit a median value of fractal 
dimension similar to that in a 3-dimensional Anderson model ~\cite{Giordano:2021qav}.

To summarize, all eigenvalues of the Dirac operator we have measured are below the 
magnetic scale $g^2T/\pi$~\cite{Linde:1980ts} and are inherently non-perturbative 
in nature. Away from the crossover transition, these (bulk) eigenmodes are delocalized 
almost over the entire volume albeit with local thermal fluctuations, consistent with RMT. 
However approaching the crossover transition results in the appearance 
of the fractal-like intermediate eigenmodes, which carry information about the chiral 
phase transition, which could be of $O(4)$ universality 
~\cite{HotQCD:2019xnw, Kaczmarek:2021ser,Kaczmarek:2023bxb}. These eigenmodes are prominently 
detectable in a temperature range where there are proposals of an IR conformal phase~\cite{Alexandru:2015fxa,Alexandru:2019gdm,Alexandru:2021pap, Horvath:2021zjk,Meng:2023nxf,Alexandru:2023xho,Alexandru:2024tel}, appearance of emergent 
symmetries~\cite{Glozman:2022lda,Rohrhofer:2019qwq} 
and their eventual disappearance can explain the \emph{effective} restoration of $U_A(1)$ 
subgroup of chiral symmetry~\cite{Kaczmarek:2023bxb}. The occurrence of percolating clusters of 
Polyakov loop~\cite{Satz:1998kg,Forrester:1993vtx}, has been reported earlier in another 
context, near the deconfinement phase transition in gauge theories without dynamical quarks. 
Furthermore, the connection between phase transitions and fractals has also been discussed  
in the context of spin systems, e.g., in the Ising model~\cite{10.1143/PTP.69.65}. 

\textbf{Properties of a non-equilibrium state of SU(3) }
Since the thermal state in QCD at high temperatures exhibits spectral 
properties corresponding to a random matrix theory, we next investigate 
if there is a classical state that exhibits chaos. Inspired by the Color Glass 
Condensate effective theory description of QCD~\cite{Gelis:2010nm}, where the typical gluon momenta 
are close to the saturation scale $Q$, which satisfies the condition $Q^2\gg \Lambda_{QCD}^2$, 
we start with an initial classical state of the non-Abelian SU(3) gauge theory 
characterized by the phase-space distribution 
$g^2f_g(|\mathbf{p}|,t=0)=n_0\frac{Q}{|\mathbf{p}|} e^{-\frac{|\mathbf{p}|^2}{2Q^2}}$. 
This initial state is clearly athermal and represents a typical over-occupied infrared 
sector of QCD, characterized by classical gluons whose occupation numbers $\sim n_0/g^2$ 
are non-perturbatively large (since $g$ is very small)~\cite{Berges:2013eia}. Starting with this initial 
condition, the gauge links $U_{i,\xs}$ and its conjugate momenta i.e. electric fields 
$E^i_{a,\xs}$ are evolved in time according to classical Hamiltonian equations of motion 
denoted by 
\begin{equation}
\frac{\partial U_{i,\xs}}{\partial t}=\frac{\partial H}{\partial E_{i,\xs}}~,~
\frac{\partial E_{i,\xs}}{\partial t}=-\frac{\partial H}{\partial U_{i,\xs}}~,~i=1,2,3~.
\end{equation}
Hamilton's equations are written in the temporal-axial gauge $A_0(\xs)=0$. 
It is well known that such a system undergoes a rapid memory loss~\cite{Berges:2013fga,Epelbaum:2013ekf} 
of their initial conditions and subsequently enter a self-similar scaling regime where 
the gluon distribution function exhibits a scaling relation of the form,
$g^2 f_g(|\mathbf{p}|,t)=(Qt)^{-\frac{4}{7}}~f_s\left[(Qt)^{-\frac{1}{7}}\frac{|\mathbf{p}|}{Q}\right],$
characteristic of a non-thermal fixed point of the classical evolution. Within such a 
scaling regime, one observes a separation of scales~\cite{Berges:2023sbs} where the 
 hard (ultraviolet) scale is distinctly separated from the electric and the 
(deep-infrared) magnetic scales similar to what is observed in equilibrium gauge theories
at high temperatures. How such a system achieves thermal equilibrium 
is not well understood, though kinetic theories that share similar 
scaling properties determined by the same critical exponents can be 
identified~\cite{Berges:2013eia,Kurkela:2015qoa}. 

\begin{figure}[h!]
    \centering
    \raisebox{-\height}{\includegraphics[width=0.4\textwidth]{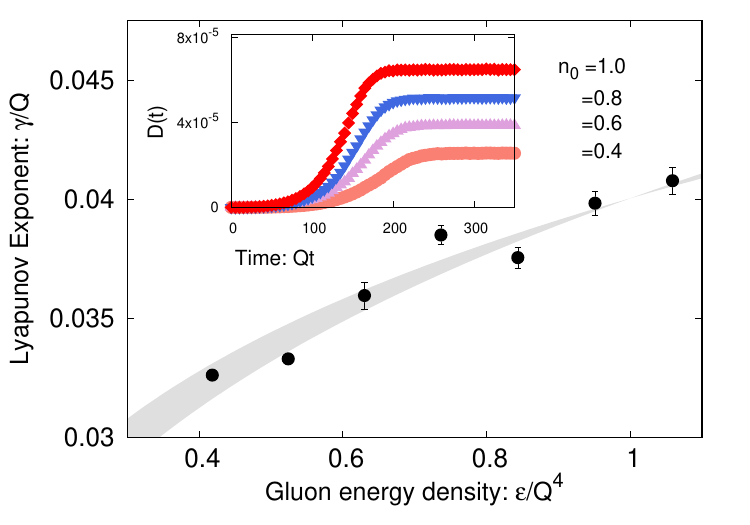}}
    \caption{Scaling of the Lyapunov exponent $\gamma$ for SU(3) with the energy density $\varepsilon$. 
    The exponential growth of the distance function $D(t)$ between two classical trajectories in the 
    gauge space for different values of initial gluon density is shown in the inset. 
    %We extract $\gamma$ by fitting the exponentially rising part with the ansatz $D(t)=D_0~\mathrm{exp}(\gamma t)$.
    }
    \label{fig:n0_vs_LyapunovExp}
\end{figure}

We next study the properties of the non-Abelian gauge theory within this self-similar 
scaling regime in particular, whether it exhibits chaotic behavior. We study the separation 
of gauge trajectories in the phase space characterized by a gauge-invariant distance 
measure~\cite{Muller:1992iw} defined as
$D(U_l, U_l^{'},t)=\frac{1}{N_P}\sum_P \frac{1}{N_c} \vert \text{tr}U_P-\text{tr}U_P^{'}\vert $,
in terms of the difference between the expectation values of the plaquettes $U_P$ and $U_P^{'}$ 
measured a time $t$ starting from infinitesimally close initial conditions at $t=0$ given 
by $n_0$ and $n_0+\Delta n_0,~\Delta n_0=0.001$. The trace is performed over 
the color degrees of freedom. The results obtained for the distance 
function are shown in the inset of Fig.~\ref{fig:n0_vs_LyapunovExp}. 
It is evident that the system shows a chaotic behavior where $D(t)$ increases exponentially 
with time, saturating at later times due to finite gauge-space volume. Fitting this initial growth 
as a function of time with the ansatz, $D(t)=D_0~\mathrm{exp}(\gamma t)$ we obtain a 
positive (largest) Lyapunov exponent $\gamma$ characteristic of a chaotic system for a wide range 
of initial gluon energy densities $\varepsilon/Q^4$ shown in Fig.~\ref{fig:n0_vs_LyapunovExp}, 
given by $\varepsilon \propto n_0/g^2$. For the range of energy densities, $\gamma/Q$ grows 
linearly as the fourth root of energy density with the best fit given by 
\begin{equation}
\gamma/Q= 0.0400(7) \times \left(\varepsilon/Q^4\right)^{0.24(3)}~.    
\end{equation}
Since the Lyapunov exponent characterizes how fast the gluon trajectories will spread out 
over the entire phase space, its inverse can give a rough estimate of the typical thermalization 
time~\cite{Gong:1992yu,Biro:1994sh}.

\textbf{Implications of our study for thermalization of non-Abelian gauge theories}
Establishing that a particular classical state of SU(3) gauge theory which 
exhibits a non-thermal scaling property in its phase-space distribution, is chaotic 
we now discuss its implications for thermalization. Within this self-similar scaling 
regime, the magnetic ($\sqrt\sigma$), electric ($m_D$) and the hard ($\Lambda$) scales 
evolve with time according to the following relations: 
$\sqrt\sigma (t) \sim Q (Qt)^{-3/10}$, $m_D (t) \sim Q (Qt)^{-1/7}$ and $\Lambda (t)\sim Q (Qt)^{1/7}$~\cite{Berges:2017igc}.
On the other hand, for SU(3) gauge theory at thermal equilibrium at $T\simeq 624$ MeV, the magnetic, 
electric, and hard scales are $g^2T/\pi=0.81T$, $gT=1.61T$ and $\pi T$ respectively. The strong coupling 
$g$ in this work is measured at the scale $\pi T$ using the five-loop $\beta$-function~\cite{Baikov:2016tgj}. 
In order to estimate the thermalization time starting from a self-similar scaling regime and ending 
at a thermal fixed point, we measure how long it takes for the magnetic scale $\sqrt\sigma (t)$ 
to evolve to its value in a thermal plasma. This is motivated by the fact that the gauge momentum modes whose 
momenta are less than the magnetic scale are over-occupied and exhibit chaotic behavior 
both in the classical as well as in the quantum (thermal) regime.  If one assumes that thermalization 
is achieved at a typical temperature $T\simeq 624$ MeV, the evolution of the magnetic scale to its thermal 
value denoted by $Q(Q\tau_{th})^{-3/10}= 0.81 T$ gives an estimate of the thermalization time 
$\tau_{th} \approx 5.2$ fm/c. Incidentally, this estimate is close to that obtained from the 
inverse of Lyapunov exponent $\gamma^{-1}\sim 4.9(1)$ fm/c measured at a typical energy density 
$\varepsilon= 0.209~Q^4$ which is closest in magnitude to the thermal plasma at $T\simeq 624$ 
MeV and also close to the estimates obtained from entropy production arguments~\cite{Muller:1992iw,Kunihiro:2010tg}.

Interestingly, though, the presence of dynamical fermions does not affect the universality in 
spectral properties of QCD below the magnetic scale (see Fig.~\ref{fig:average_rtilda}) or 
quantities like the spatial string tension $\sqrt{\sigma}$, even in the non-thermal state. 
This comes from the fact that $\sqrt\sigma$ only increases by $\sim 1.6\%$ in the presence of dynamical 
quarks~\cite{Bala:2025ilf} in QCD at thermal equilibrium for $T> 600$ MeV. Fermion production is 
an inherent quantum process that will pull the system out from the athermal classical fixed point 
to the thermal one and would facilitate fast thermalization. This can be understood from the fact 
that the presence of dynamical quarks increases the magnitude of the magnetic scale in a thermal 
system by modifying the running of the gauge coupling with energy. At temperature $T\simeq 624$ MeV, 
the presence of dynamical quarks leads to an enhancement in the magnetic scale to $g^2T/\pi=1.19T$ 
which is $\sim 1.5$ times compared to its value in gauge theory without fermions. 
The effect on the thermalization time is dramatic since it varies with the ratio of magnetic scales 
as $(0.81/1.19)^{10/3}$. The fermions thus facilitate a faster rate of thermalization, resulting in 
a $\tau_{th}=1.44$ fm/c which is about $\sim 28\%$ of that estimated earlier in gauge 
theory without fermions. This upper bound estimate of $\tau_{th}$ reported here is smaller compared to 
the typical thermalization time-scales $\sim 3$ fm/c estimated using perturbative QCD scattering 
processes~\cite{Heinz:2004pj}.

\textbf{Conclusions \& Outlook}
Over the years, the features inherent in the Dirac eigenspectrum have provided 
us with key insights on the chiral, topological and localization properties of 
QCD~\cite{Sharma:2018y2,Lombardo:2020bvn,Giordano:2021qav}. In this Letter, we discuss 
two interesting aspects of the Dirac spectrum of QCD and their physical implications. 
Firstly, in QCD with $2+1$ flavors of physical quarks, extended eigenvectors percolating 
over the entire volume start to become prominent just above $T_c$ which show intermediate 
level statistics between opposite extremes of an RMT and uncorrelated eigenvalues. Though 
the transition associated with chiral symmetry breaking is a smooth crossover and hence there 
is no unique order parameter, however the structural features of the eigenvectors close to $T_c$ 
are observed to carry information about the universality class. This is evident from the fact that 
the median values of the fractal dimensions of these intermediate eigenmodes are consistent with $O(4)$ 
universality. Secondly, we demonstrate a non-trivial realization of the BGS conjecture in 
non-Abelian SU(3) gauge theory, by measuring its single-particle states with the Dirac operator defined on 
these gauge links. Whereas eigenstates of the Dirac operator measured on a thermal (quantum) QCD configuration 
which are deep in the bulk, but with eigenvalues below the magnetic scale $g^2T/\pi$, exhibit level-spacing ratios 
consistent with random matrices of the Gaussian unitary ensemble, there exists a classical non-thermal state 
of magnetic gluons which is chaotic.  Establishing this fact allows us to naturally ask the follow-up question: 
How long does it take for the chaotic system consisting of magnetic gluons to thermalize?

Defining thermalization time as how fast these soft single-particle momentum states acquire a
thermal distribution starting from an over-occupied state of gluons where single-particle momentum
distribution exhibits a self-similar scaling, we obtain an upper bound of $\sim 1.44$ fm/c 
to reach to a thermal state at $T\simeq 624$ MeV by matching the (magnetic) scales, 
below which these states are defined. The inclusion of dynamical 
fermions only results in an efficient separation of scales at these high temperatures, 
thus allowing for such a fast thermalization of the non-perturbative but classical 
infrared (magnetic) modes of QCD. It would nonetheless be interesting to verify this 
picture with quantum simulations of dynamical quarks with gauge interactions which could 
be feasible in lower dimensions in the coming years.

\textbf{Acknowledgements}
We are grateful to Soumya Bera, Sumilan Banerjee, Frithjof Karsch, V. K. B. Kota, Hridis K. Pal 
and Soeren Schlichting for helpful suggestions and discussions during the course of this work. We are 
indebted to Frithjof Karsch and Soeren Schlichting for their kind hospitality at the Theoretical 
Physics Group, Bielefeld University when this work was being finalized. This work is in part 
supported by the Deutsche Forschungsgemeinschaft (DFG, German Research Foundation) - Project number 
315477589-TRR211.  We thank the HotQCD collaboration, formerly also consisting of
members from the RBC-LLNL collaboration, for sharing some of the gauge configurations with us. The 
authors acknowledge the computing time provided by the Computing Center at the Institute of 
Mathematical Sciences. Our GPU code is in part based on some of the publicly available 
QUDA libraries~\cite{Clark:2009wm}.

\bibliography{ArXiv.bib}

\end{document}